\documentclass[letterpaper]{article}
\usepackage{iccc}

\usepackage{color}
\usepackage{times}
\usepackage{helvet}
\usepackage{courier}
\usepackage{graphicx}
\usepackage{subcaption}
\usepackage{caption}
\usepackage{amsmath}
\usepackage{url}
\usepackage{booktabs}
\usepackage{multirow}

\title{Dance2Music: Automatic Dance-driven Music Generation}

{
\author{
    Gunjan Aggarwal$^{1}$ \hspace{1.5pc}
    Devi Parikh$^{2}$ \\
    $^1$Adobe,~~
	$^2$Facebook AI Research \& Georgia Tech\\
	$^1${\tt guaggarw@adobe.com},~~
	$^2${\tt parikh@gatech.edu}
	}
}

\begin{document} 
\maketitle
\begin{abstract}
\begin{quote}
Dance and music typically go hand in hand. The complexities in dance, music, and their synchronisation make them fascinating to study from a computational creativity perspective. While several works have looked at generating dance for a given music, automatically generating music for a given dance remains under-explored. This capability could have several creative expression and entertainment applications. We present some early explorations in this direction. We present a search-based \emph{offline} approach that generates music after processing the entire dance video and an \emph{online} approach that uses a deep neural network to generate music on-the-fly as the video proceeds. We compare these approaches to a strong heuristic baseline via human studies and present our findings. We have integrated our \emph{online} approach in a live demo! A video of the demo can be found here: \url{https://sites.google.com/view/dance2music/live-demo}.

\end{quote}
\end{abstract}

\section{Introduction}
American poet Henry Wadsworth Longfellow called music the universal language of mankind. ``Music is the art of arranging sounds in time to produce a composition through the elements of melody, harmony, rhythm, and timbre" (Wikipedia). Given this richness, researchers have extensively explored automatic music generation. Jukebox~\cite{dhariwal2020jukebox} is 
a recent work that generates music of different genres, artists, and lyrics.

Music is naturally connected with dance. It sets the mood, maintain a dancer's beat and can amplify a dance's emotional affect. In fact, music and movement are represented similarly in the brain, using a shared neural code \cite{Sievers254961}. This connection between dance and music makes it interesting to study the two complimentary art forms together. While several deep learning~\cite{huang2020dance,lee2019dancing} and search~\cite{tendulkar2020feel} based solutions have been proposed to generate dance for a given music, the other direction of automatically generating music for a dance is still relatively unexplored. 

Such an algorithm has applications in creating short-form videos popular on social media platforms, in home assistants that can generate music as a family is playfully dancing around, in adding rhythm to our workouts making them more enjoyable, etc. By allowing people to generate music via their body movements, the tool enables an interesting confluence of our senses.

\begin{figure}
    \centering
    \includegraphics[width=0.9\columnwidth]{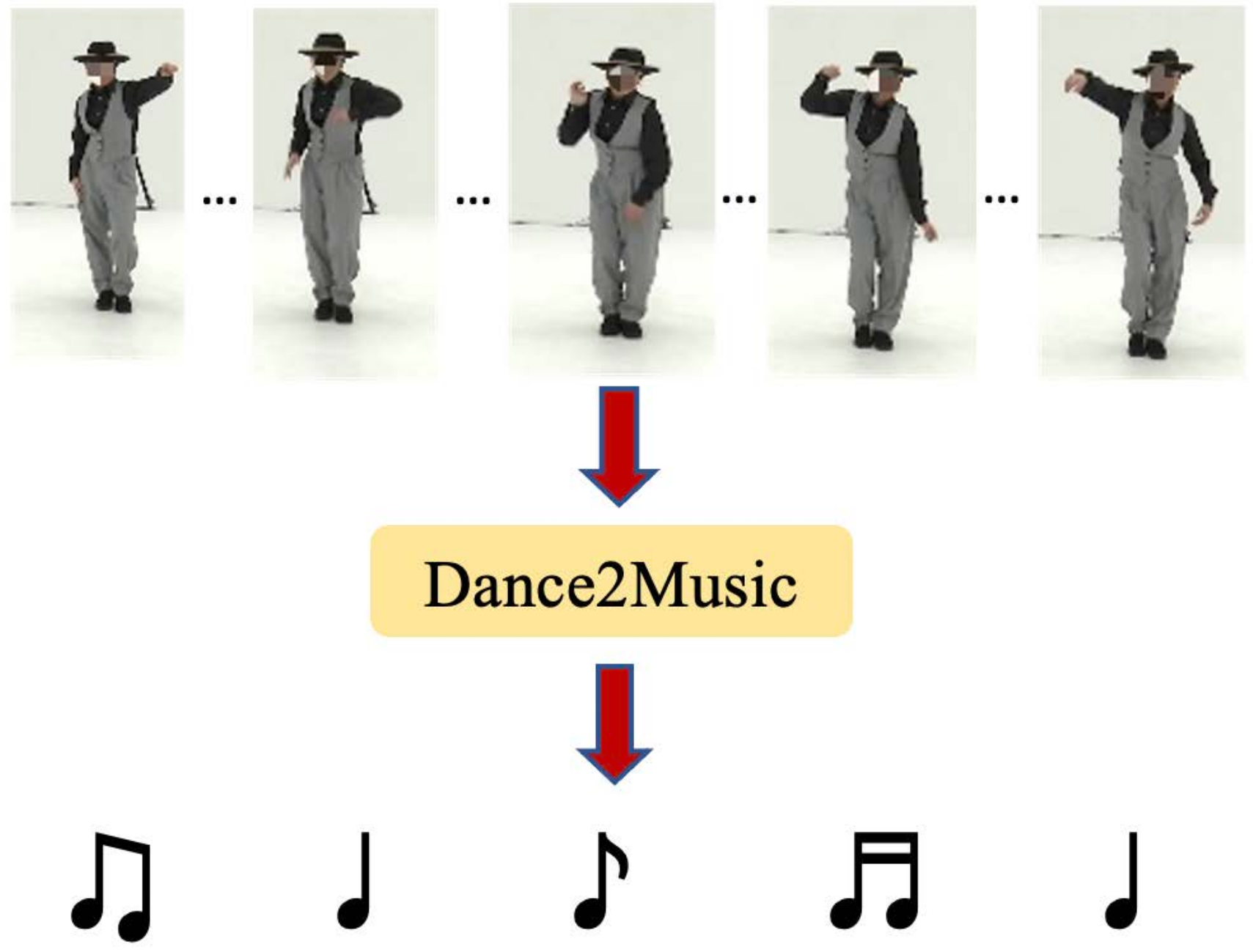}
    \caption{Generating music from dance.}
    \label{fig:my_label}
    \vspace{-10pt}
\end{figure}

As a preliminary step in this direction, we represent music as a sequence of notes from the C major pentatonic scale. Dance is processed as a series of human poses extracted from an existing video or an incoming video feed. We generate a sequence of notes that aligns with the movement pattern in the dance poses. We present a search-based \emph{offline} approach that generates music after processing the entire dance video and an \emph{online} approach that uses a deep neural network to generate music in real-time.
We have integrated this \emph{online} approach in a live demo. A video of the demo can be found here: \url{https://sites.google.com/view/dance2music/live-demo}. We also present a strong baseline and compare our approaches on 10 dance clips via human studies. We also quantitatively evaluate our \emph{online} approach on 45 videos.

\section{Related Work}
\noindent \textbf{Generating dance from music.} 
Several works exist to automatically generate dance from music. The inverse is relatively under explored. Early works~\cite{5753889}~\cite{lee2013music} generate dance from music using retrieval-based methods. \cite{tendulkar2020feel} propose a search based approach to generate dance from music such that the overall spatio-temporal movement pattern matches the overall structure of music. Several deep learning approaches~\cite{lee2019dancing}, \cite{huang2020dance} have also been proposed to do the same.\\
\noindent \textbf{Generating sound from silent videos.} Several approaches reproduce missing sound from a video clip of people playing musical instruments
~\cite{gan2020foley} or someone sneezing or coughing ~\cite{chen2020generating}. Our task is to generate novel music that goes well with the dance, rather than reproduce reality. In fact, there may have been no music playing with the dance in reality.\\
\noindent \textbf{Controlling audio from body movement.} 
Several works use sensors to control music generation through hand movements and eye gaze~\cite{nawaz2008infotainment},~\cite{vaidya2019hand}.
~\cite{10.2307/3681529} track movement by attaching sensors to bodies of dancers  to guide music generation. In contrast, our setup has a lower barrier to entry (it uses just a camera) and can also generate music post hoc (from a dance video).

\section{Approach}
Our goal is to generate music that goes well with an input dance. Similar to \cite{tendulkar2020feel}, we hypothesize that if notes are (dis)similar when the associated poses in the dance are (dis)similar, the music will feel synced with the dance.
Our approach maximizes this correlation between poses and notes.\\
\noindent \textbf{Dance representation.} 
In this preliminary work, we assume a single dancer in the video. We represent dance as a sequence of poses. For an existing dance video, we extract 30 frames per second and use OpenPose~\cite{cao2019openpose} to estimate the person's pose for each extracted frame. Each pose consists of 18 2D keypoints\footnote{https://github.com/CMU-Perceptual-Computing-Lab/openpose/blob/master/doc/02\_output.md\#pose-output-format-coco}
and is represented by a 36-dimensional continuous vector normalized by the image size so that the values are between -1 and 1. The same representation is used when processing a live dance video feed.\footnote{To run a live demo on CPU we use Intel OpenVINO toolkit~\cite{intel} because of its optimized inference on CPU.}
\noindent \textbf{Music representation.}
We use piano notes in the C major pentatonic scale -- C4, D4, E4, G4 and A4 -- to compose our music. The pentatonic scale is found in different forms in most of the world's music~\cite{penta}. Additionally, notes in a pentatonic scale are so consonant with each other that even a random sequence is pleasing to listen to. 

\begin{figure}[t]
\centering
\begin{subfigure}[b]{0.15\textwidth}
    \includegraphics[width=\textwidth]{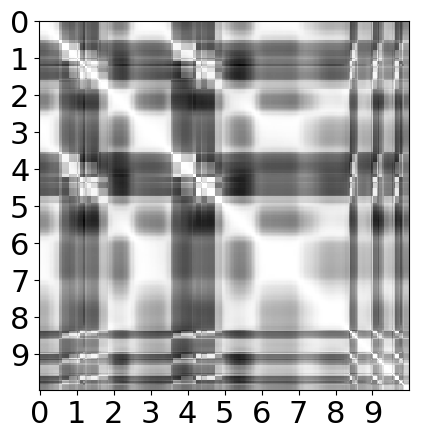}
    \caption{\scriptsize{Dance similarity matrix.}}
\end{subfigure}%
~
\begin{subfigure}[b]{0.15\textwidth}
    \includegraphics[width=\textwidth]{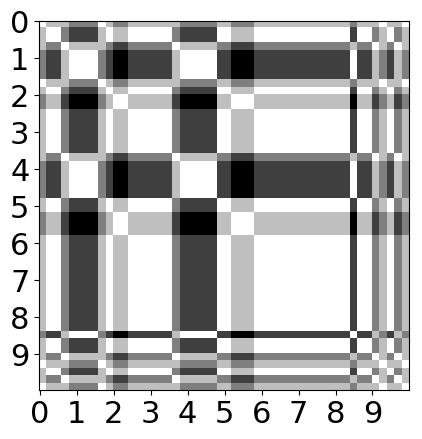}
    \caption{\scriptsize{Music similarity matrix using global history.}}
\end{subfigure}
~
\begin{subfigure}[b]{0.15\textwidth}
    \includegraphics[width=\textwidth]{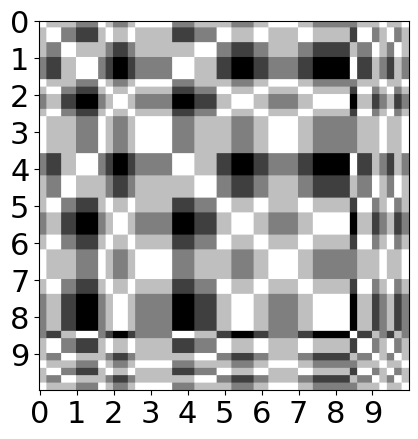}
    \caption{\scriptsize{Music similarity matrix using local history.}}
\end{subfigure}
\caption{Example dance and music similarity matrices for our approaches. The axis labels indicate the time (in secs). Note that the music produced by using global history (b) looks flat (identical notes played for long duration of time) compared to using local history (c).}
\label{fig:local_global}
\vspace{-10pt}
\end{figure}

\begin{figure}[t]
    \centering
    \includegraphics[width=\columnwidth]{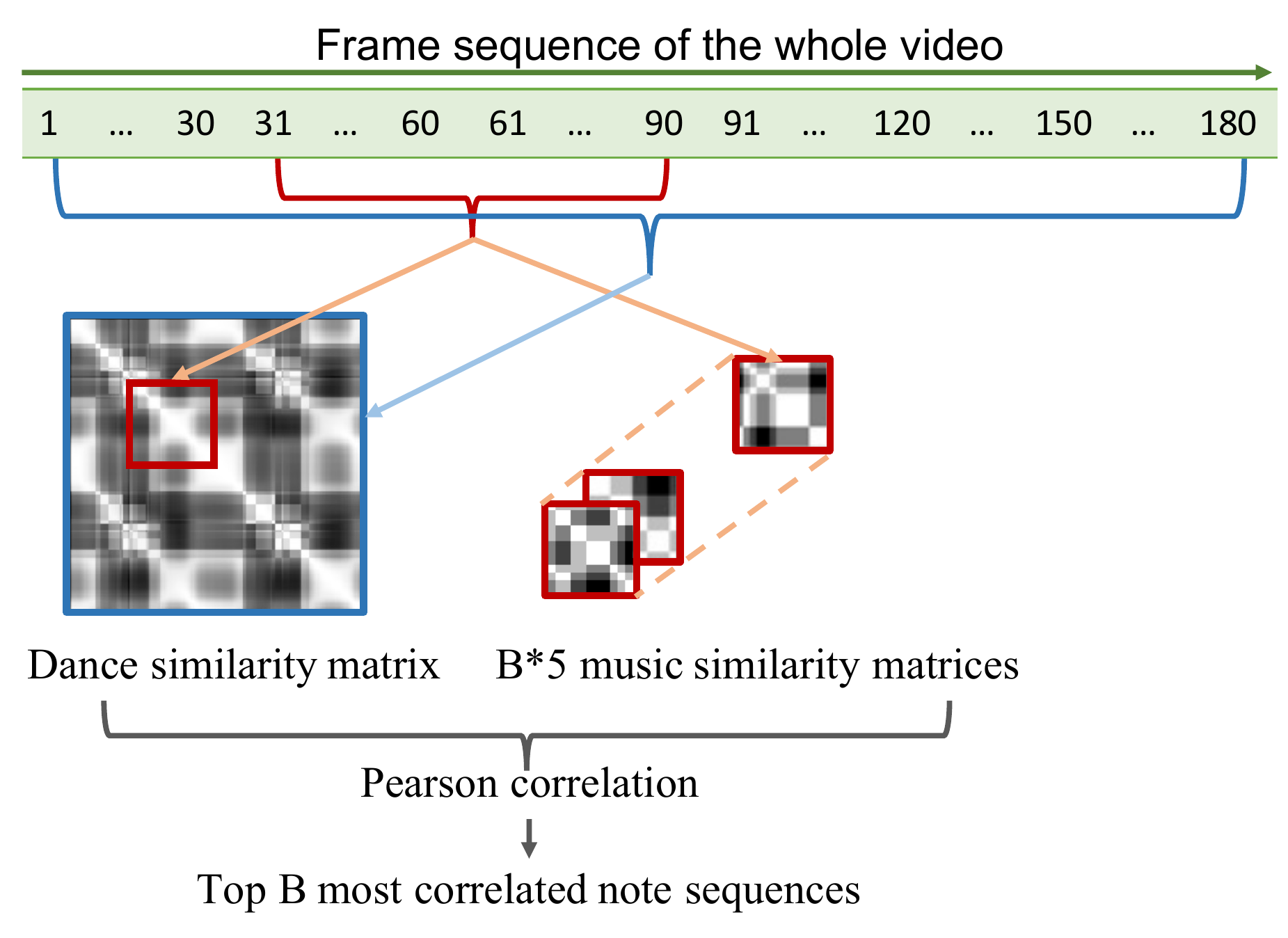}
    \caption{An overview of our \emph{offline} music generation approach. At each time step, we look at the dance similarity matrix corresponding to a past local window (highlighted in red) and sort each of the $B*5$ local candidate note sequences using Pearson correlation. The top $B$ correlated sequences are kept for the next steps of beam search, and the process repeats till notes have been generated for the entire video.
    }
    \label{fig:flowDig}
    \vspace{-10pt}
\end{figure}

\subsection{Offline Approach}
In our offline approach, we generate music for an existing dance video. Given a sequence of $N$ dance frames, and a hyper-parameter $K$, we generate a music sequence of length $\frac{N}{K}$. $K$ -- the \emph{music interval} -- denotes the number of dance frames after which we play a musical note. Based on an initial qualitative assessment, in our experiments we set $K$ to be 6. Recall that our aim is generate music, such that the generated music is maximally correlated with the dance.

\noindent \textbf{Correlation.}
We bring dance and music to a unified similarity matrix representation and then compute the Pearson correlation between the two matrices vectorized.
Let $D$ be the dance similarity matrix, and $M$ be the music similarity matrix. We define them as follows. \\
$$D[i][j] = \text{cosine\_similarity}(\text{pose[i]}, \text{pose[j]})$$
$$M[i][j] = \frac{\text{abs(note[i]-note[j])}}{4.0}$$

\noindent Here $\text{pose[i]}$ denotes the 36 dimensional representation for the $i^{th}$ dance frame, and $\text{note[i]}$ represents the note played at $i^{th}$ music interval. We represent the notes C4, D4, E4, G4 and A4 in an ordinal manner as 0, 1, 2, 3 and 4, ordered by frequency. The matrices capture how similar the two poses or notes are in two points in time. In order to compute correlation between the two, we resize $M$ to be of the same size as $D$ using nearest neighbour interpolation. Recall that $M$ is smaller than $D$ because we only play a note every $K$ frames.

Our objective is to choose a sequence of notes such that the Pearson correlation between the corresponding music and dance similarity matrix is maximized. As the search space is exponential in the number of notes (e.g., $5^{60}$ for a 12-second video), an exhaustive search is infeasible.

\noindent \textbf{Search.} We use beam search to find the sequence with the highest correlation with the dance. With the first note fixed at E4, the generation advances from left (start of the video) to right (end of the video). At each step of beam search, we maintain a candidate list of $B$ sequences of notes that currently have the highest correlation with the dance. We append each of the 5 candidate notes to each sequence, resulting in $B\times5$ sequences. Each of these sequences are scored using the correlation with the dance. We keep the top $B$, and discard the rest. This is repeated till $\frac{N}{K}$ notes have been generated. Instead of using the entire history of dance (and notes) in computing the correlation, we use the local history (of 10 notes and associated 60 frames). We find that using the local history allows us to capture more minute details in the dance, while using the global history results in more ``flat'' music (see Figure~\ref{fig:local_global}). This approach is illustrated in Figure~\ref{fig:flowDig}. We use $B=50$ in our experiments. 

\subsection{Online Approach}
The above approach cannot be run in real-time due to beam search -- the note to be produced at a given time $t$ is not decided until the computation for future notes has been done. To enable real-time note generation, we model the problem using neural networks. We run the \emph{offline} approach on a set of videos to collect paired dance and music data. Using this data, we train a neural network to take in as input the local history of the dance poses and generated notes, and produce as output the note to be played next. We model this as a 5-way classification task (to generate one of the 5 notes).

\begin{figure}[!h]
    \centering
    \includegraphics[width=\columnwidth]{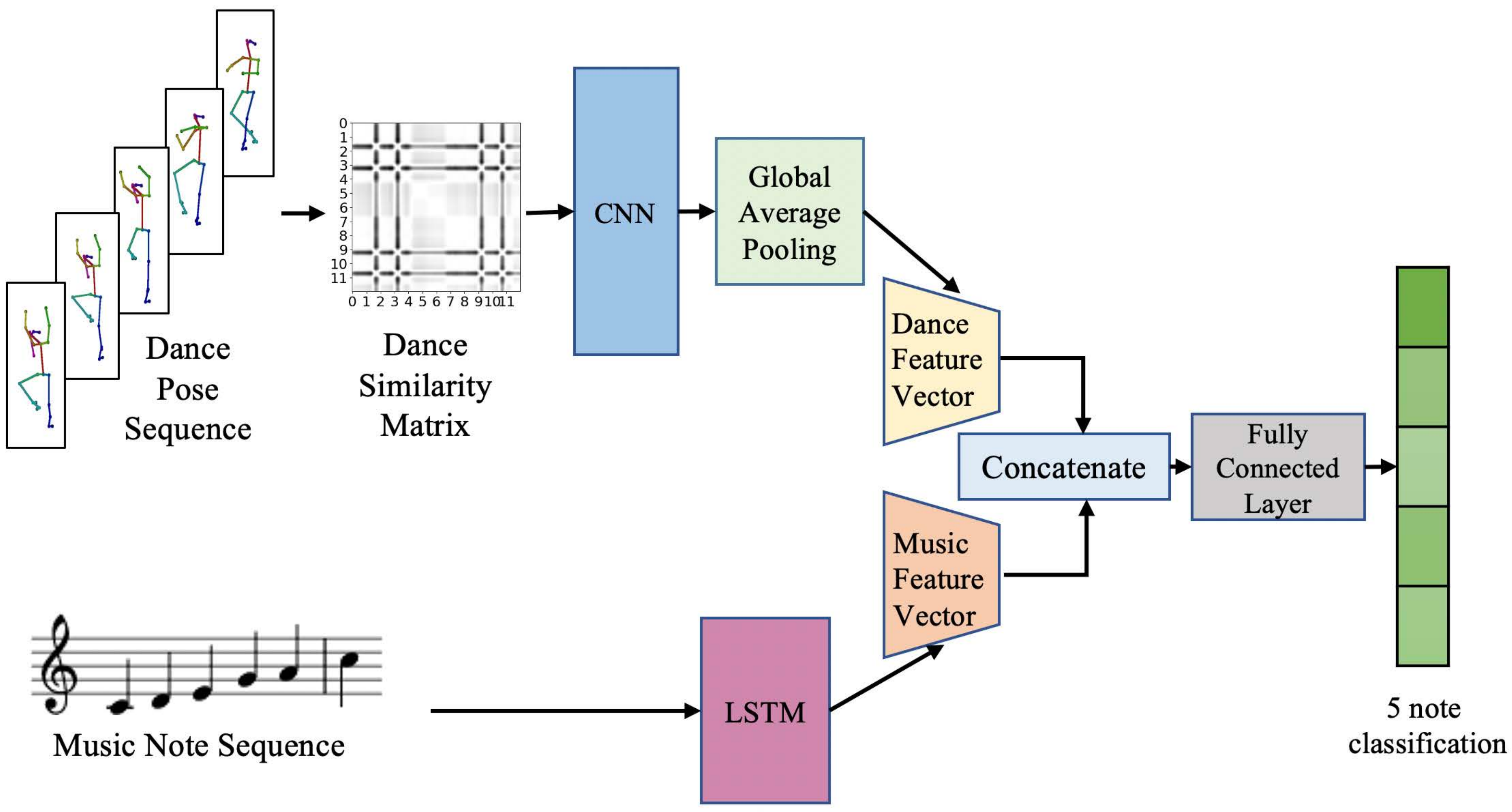}
    \caption{
Neural network architecture for note prediction. Our network gets as input the local history of both dance similarity matrix and note sequences up to time $t$ and predicts the note for time $t+1$.}
    \label{fig:nn_architecture}
    \vspace{-10pt}
\end{figure}

Figure~\ref{fig:nn_architecture} shows the model architecture. We process the past dance similarity matrix via 6 CNN layers, each of kernel size 3 and filters 64, 128, 128, 256, 512 and 32 respectively. Max pooling is done after the first and third convolution and ReLU activation is used after each CNN layer. Global average pooling is performed to get the dance feature representation. The past music note sequence is fed as an input to an LSTM with hidden dimension 32 and the resulting features are concatenated with the dance features. Three fully connected (FC) layers of sizes 512, 256 and 128, each followed by ReLU activation are applied over the concatenated features. Finally an FC layer of size 5 is applied to get the final logits. The network is trained with Adam optimizer for 200 epochs with a learning rate of 2e-4. The dance similarity matrix and the music sequence are padded to ensure that each data point in a batch is of the same size.

For generating music on-the-fly as dance frames come in, we start with the first note being E4, and then sample iteratively from our trained model. The note predicted at time $t$ along with local history of both generated notes and dance poses is fed as an input for the note prediction at time $t+1$.  

The task that the neural network is modelling is difficult because it has access to partial information. For the \emph{offline} approach, the note generated at any time is determined based on not just past but also future notes and frames. However, our neural network doesn't have access to future notes and frames (thus enabling the ``online'' use case). 

\vspace{3pt}
\noindent \textbf{Baseline}
Inspired by \cite{boden92} that creativity is a combination of quality (value) and surprise (novelty), we design a baseline that generates a sequence of notes that is both in sync with the dance and unpredictable. We select a random note when the similarity between poses across a music interval is below a threshold, otherwise continue playing the previous note. We use the $80^{th}$ percentile of all similarities over the training dance frames as the threshold. We choose $80^{th}$ percentile instead of median to ensure that a note change occurs frequently and the generated music is not flat. 

Examples of music composed by our various algorithms can be found here: \url{https://sites.google.com/view/dance2music}.

\section{Results}
\noindent \textbf{Dataset}
We perform our experiments using the AIST dataset~\cite{tsuchida2019aist} and videos from~\cite{chan2019dance}. The former consists of street dance performed by different people across 10 different genres. It has each dance video captured from 9 angles; we use the ones with front facing camera view. The latter consists of short YouTube videos where a single subject dances in front of a static camera. All videos are ${\sim}12$ seconds long. 

\begin{figure}[t]
\centering
\begin{subfigure}[b]{0.15\textwidth}
    \includegraphics[width=\textwidth]{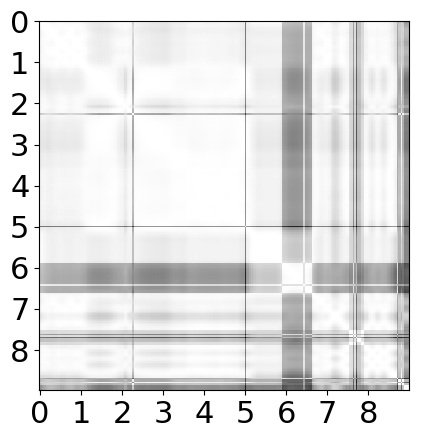}
    \caption{\scriptsize{Dance similarity matrix.}}
\end{subfigure}%
~
\begin{subfigure}[b]{0.15\textwidth}
    \includegraphics[width=\textwidth]{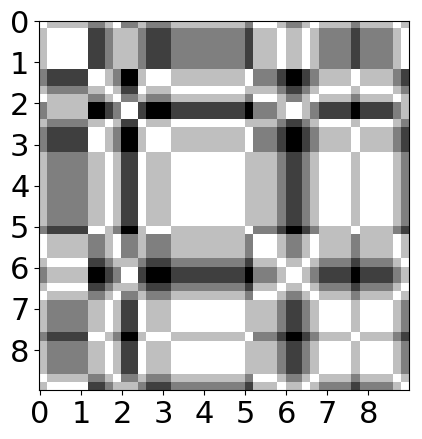}
    \caption{\scriptsize{Music similarity matrix for offline approach.}}
\end{subfigure}
~
\begin{subfigure}[b]{0.15\textwidth}
    \includegraphics[width=\textwidth]{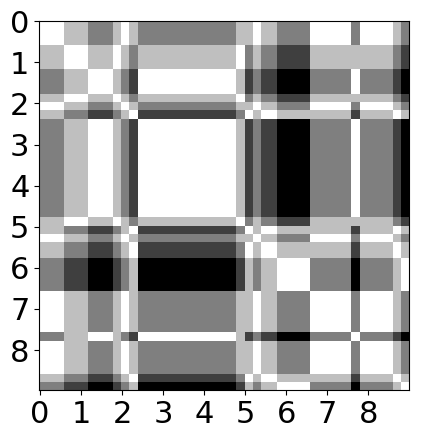}
    \caption{\scriptsize{Music similarity matrix for online approach.}}
\end{subfigure}
\caption{Example dance and music similarity matrices for our approaches. The axis labels indicate the time (in secs).}
\label{fig:offline_online}
\end{figure}

\noindent \textbf{Automatic metrics.} We start by evaluating how well the neural network in our \emph{online} approach mimics the \emph{offline} approach. We train the model on 455 videos from the AIST dataset, and evaluate it on 45 videos. It achieves a test accuracy of 73.5\% at the task of predicting the next note accurately (that is, match the note produced by the \emph{offline} approach). Note that chance performance would be 20\%. We also compute the Pearson correlation between the generated note sequence and the dance sequence -- the metric the \emph{offline} approach was optimizing for -- across all the test videos. The mean correlation of the \emph{online} model is 0.33. As a reference, the mean correlation of the \emph{offline} approach (which our neural network model is mimicking) is 0.38. Figure~\ref{fig:offline_online} shows dance and music similarity matrices for both approaches.

\noindent \textbf{Human Study.}
Next we conduct human studies to compare our approaches to the strong baseline presented earlier. We use 10 videos -- 8 from AIST and 2 from~\cite{chan2019dance}. This gives us 10 video pairs for each comparison type: \emph{offline} vs. baseline and \emph{online} vs. baseline. For each comparison we show subjects a pair of videos, both contain the same dance but the music is generated by the different approaches. We ask subjects: \emph{Which music composition goes better with the dance?} Each pair is evaluated by 3 subjects. 18 subjects (11 male, 7 female) voluntarily participated in the study. Their ages range from 16 to 49 years. Each subject evaluated up to 3 video pairs for each comparison type.

A one-sample proportion hypothesis test suggests that for our sample size of 30 (3 subjects $\times$ 10 video pairs), a “win ratio” over 0.633 (19 out of 30), or below 0.367 (11 out of 30) is statistically significant at 95\% confidence.

Table~\ref{table:results} shows the results of our human study. For \emph{offline} vs. baseline, \emph{offline} was preferred 77\% of the times (23 out of 30), which is statistically significant. This shows that our search based approach that finds a sequence of notes so that (dis)similar poses in the dance have (dis)similar notes associated with them generates music that subjects find goes better with the dance than a strong baseline that includes both sync (quality) and randomness (novelty).

Subjects preferred the \emph{online} approach over the baseline 70\% of the time (21 out of 30), which is also statistically significant. This, combined with the automatic metrics reported above where the \emph{online} approach mimics the \emph{offline} approach well, suggests that our neural network model is a promising direction for generating music on-the-fly that goes well with a live dance. A video of a live demo of our \emph{online} approach can be found here \url{https://sites.google.com/view/dance2music/live-demo}.

\begin{table}
\begin{tabular}{@{}l|l|l|l@{}}
\toprule
Comparisons                           & Approach & Score & Win Ratio \\ \midrule
\multirow{2}{*}{Offline vs. Baseline} & Offline  & 23    & 0.77      \\ \cmidrule(l){2-4} 
                                      & Baseline & 7     & 0.23      \\ \midrule
\multirow{2}{*}{Online vs. Baseline}  & Online   & 21    & 0.70      \\ \cmidrule(l){2-4} 
                                      & Baseline & 9     & 0.30      \\ \bottomrule
\end{tabular}
\caption{Results of human evaluation performed for comparing \emph{offline} and \emph{online} approach against our baseline.}
\label{table:results}
\vspace{-10pt}
\end{table}

\section{Conclusion and Future Work}

We present a preliminary study for automatic music generation conditioned on input dance. We present two approaches, one for offline generation for an existing video, and one for real-time generation for an ongoing dance. We compare the approaches against a strong baseline using human studies. Human subjects find our \emph{offline} and \emph{online} approaches to be significantly more aligned with the dance than our baseline. Future work involves exploring more dimensions of music (increasing the number of notes and instruments, including chords, playing notes at different frequencies instead of every $K$ frames), and more varied videos (different dance forms, more than one person in a video). 

\noindent \textbf{Acknowledgment.}
Larry Zitnick, Kristin Galvin and David Kant for helpful discussions.






\bibliographystyle{iccc}
\bibliography{iccc}

\begin{thebibliography}{}

\bibitem[\protect\citeauthoryear{Boden}{1992}]{boden92}
Boden, M.
\newblock 1992.
\newblock {\em The Creative Mind}.
\newblock London: Abacus.

\bibitem[\protect\citeauthoryear{Cao \bgroup et al.\egroup
  }{2019}]{cao2019openpose}
Cao, Z.; Hidalgo, G.; Simon, T.; Wei, S.-E.; and Sheikh, Y.
\newblock 2019.
\newblock {OpenPose: Realtime Multi-Person 2D Pose Estimation Using Part
  Affinity Fields}.
\newblock {\em IEEE PAMI} 43(1).

\bibitem[\protect\citeauthoryear{Chan \bgroup et al.\egroup
  }{2019}]{chan2019dance}
Chan, C.; Ginosar, S.; Zhou, T.; and Efros, A.~A.
\newblock 2019.
\newblock {Everybody Dance Now}.
\newblock In {\em ICCV}.

\bibitem[\protect\citeauthoryear{Chen \bgroup et al.\egroup
  }{2020}]{chen2020generating}
Chen, P.; Zhang, Y.; Tan, M.; Xiao, H.; Huang, D.; and Gan, C.
\newblock 2020.
\newblock {Generating Visually Aligned Sound From Videos}.
\newblock {\em TIP}.

\bibitem[\protect\citeauthoryear{Dhariwal \bgroup et al.\egroup
  }{2020}]{dhariwal2020jukebox}
Dhariwal, P.; Jun, H.; Payne, C.; Kim, J.~W.; Radford, A.; and Sutskever, I.
\newblock 2020.
\newblock {Jukebox: A Generative Model For Music}.
\newblock {\em arXiv preprint arXiv:2005.00341}.

\bibitem[\protect\citeauthoryear{Fan, Xu, and Geng}{2012}]{5753889}
Fan, R.; Xu, S.; and Geng, W.
\newblock 2012.
\newblock {Example-Based Automatic Music-Driven Conventional Dance Motion
  Synthesis}.
\newblock {\em IEEE TVCG} 18(3).

\bibitem[\protect\citeauthoryear{Gan \bgroup et al.\egroup
  }{2020}]{gan2020foley}
Gan, C.; Huang, D.; Chen, P.; Tenenbaum, J.~B.; and Torralba, A.
\newblock 2020.
\newblock {Foley Music: Learning To Generate Music From Videos}.
\newblock In {\em ECCV 2020}.
\newblock Springer.

\bibitem[\protect\citeauthoryear{Huang \bgroup et al.\egroup
  }{2020}]{huang2020dance}
Huang, R.; Hu, H.; Wu, W.; Sawada, K.; Zhang, M.; and Jiang, D.
\newblock 2020.
\newblock {Dance Revolution: Long-Term Dance Generation With Music Via
  Curriculum Learning}.
\newblock {\em arXiv preprint arXiv:2006.06119}.

\bibitem[\protect\citeauthoryear{Intel}{2020}]{intel}
Intel.
\newblock 2020.
\newblock {OpenVINO Toolkit }.
\newblock \url{https://software.intel.com/en-us/openvino-toolkit}.

\bibitem[\protect\citeauthoryear{Lee \bgroup et al.\egroup
  }{2019}]{lee2019dancing}
Lee, H.-Y.; Yang, X.; Liu, M.-Y.; Wang, T.-C.; Lu, Y.-D.; Yang, M.-H.; and
  Kautz, J.
\newblock 2019.
\newblock {Dancing To Music}.
\newblock {\em arXiv preprint arXiv:1911.02001}.

\bibitem[\protect\citeauthoryear{Lee, Lee, and Park}{2013}]{lee2013music}
Lee, M.; Lee, K.; and Park, J.
\newblock 2013.
\newblock {Music Similarity-Based Approach To Generating Dance Motion
  Sequence}.
\newblock {\em Multimedia tools and applications} 62(3).

\bibitem[\protect\citeauthoryear{Lütkebohle}{2011}]{penta}
Lütkebohle, I.
\newblock 2011.
\newblock {Pentatonic Scale }.
\newblock \url{https://www.britannica.com/art/pentatonic-scale}.

\bibitem[\protect\citeauthoryear{Morales-Manzanares \bgroup et al.\egroup
  }{2001}]{10.2307/3681529}
Morales-Manzanares, R.; Morales, E.~F.; Dannenberg, R.; and Berger, J.
\newblock 2001.
\newblock {SICIB: An Interactive Music Composition System Using Body
  Movements}.
\newblock {\em Computer Music Journal} 25(2).

\bibitem[\protect\citeauthoryear{Nawaz, Mian, and
  Habib}{2008}]{nawaz2008infotainment}
Nawaz, T.; Mian, M.~S.; and Habib, H.~A.
\newblock 2008.
\newblock {Infotainment Devices Control By Eye Gaze And Gesture Recognition
  Fusion}.
\newblock {\em IEEE Transactions on Consumer Electronics} 54(2).

\bibitem[\protect\citeauthoryear{Sievers \bgroup et al.\egroup
  }{2019}]{Sievers254961}
Sievers, B.; Parkinson, C.; Kohler, P.~J.; Hughes, J.; Fogelson, S.~V.; and
  Wheatley, T.
\newblock 2019.
\newblock {Visual And Auditory Brain Areas Share A Neural Code For Perceived
  Emotion}.
\newblock {\em bioRxiv}.

\bibitem[\protect\citeauthoryear{Tendulkar \bgroup et al.\egroup
  }{2020}]{tendulkar2020feel}
Tendulkar, P.; Das, A.; Kembhavi, A.; and Parikh, D.
\newblock 2020.
\newblock {Feel The Music: Automatically Generating A Dance For An Input Song}.
\newblock {\em arXiv preprint arXiv:2006.11905}.

\bibitem[\protect\citeauthoryear{Tsuchida \bgroup et al.\egroup
  }{2019}]{tsuchida2019aist}
Tsuchida, S.; Fukayama, S.; Hamasaki, M.; and Goto, M.
\newblock 2019.
\newblock {AIST Dance Video Database: Multi-Genre, Multi-Dancer, And
  Multi-Camera Database For Dance Information Processing.}
\newblock In {\em ISMIR}.

\bibitem[\protect\citeauthoryear{Vaidya \bgroup et al.\egroup
  }{2019}]{vaidya2019hand}
Vaidya, O.; Jadhav, K.; Ingale, L.; and Chaudhari, R.
\newblock 2019.
\newblock {Hand Gesture Based Music Player Control In Vehicle}.
\newblock In {\em IEEE I2CT}.

\end{thebibliography}

\end{document}